\documentclass{iaus}
\usepackage{graphicx}

\title[The star capture model for fueling quasar accretion disks]{The star capture model for fueling quasar accretion disks}
\author[Gareth F. Kennedy, Jordi Miralda-Escud\'e \& Juna A. Kollmeier]
{Gareth F. Kennedy$^1$, Jordi Miralda-Escud\'e$^{2,1}$ \& Juna A. Kollmeier$^3$ }
\affiliation{$^1$ Institut de Ci\`encies del Cosmos, University of Barcelona \\ 
$^2$ Institutuci\'o Catalana de Recerca i Estudis Avan\c cats \\ 
$^3$ Carnegie Observatories, Pasadena}

\pubyear{2010}
\volume{IAUS 271}  %% insert here IAU Symposium No.
\pagerange{119--126}
\jname{Astrophysical Dynamics: From Stars to Galaxies}
\editors{Nic Brummell, Allan Sacha Brun, Mark S. Miesch \&  Y. Ponty, eds.}

\begin{document}

\newcommand{\msun}{M_{\odot}}
\newcommand{\rsun}{R_{\odot}}
\newcommand{\figsizeFour}{4.5cm}

\maketitle

\begin{abstract}
Although the powering mechanism for quasars is now widely recognized to be the accretion of matter in a geometrically thin disk, the transport of matter to the inner region of the disk where luminosity is emitted remains an unsolved question. \cite[Miralda-Escud\'e \& Kollmeier (2005)]{JMK2005} proposed a model whereby quasars are fuelled when stars are captured by the accretion disk as they plunge through the gas. Such plunging stars can then be destroyed and deliver their mass to the accretion disk.

Here we present the first detailed calculations for the capture of stars originating far from the accretion disk near the zone of influence of the central black hole. In particular we examine the effect of adding a perturbing mass to a fixed stellar cusp potential on bringing stars into the accretion disk where they can be captured. The work presented here will be discussed in detail in an upcoming publication \cite{Kennedy_etal10}.

\end{abstract}

\firstsection % if your document starts with a section,
\section{Model and results}

% initial conditions
The model consists of a stellar cluster and an accretion disk both centered on a $10^{8}\msun$ massive black hole (MBH). To examine the effect of the disk on the orbits of stars at the zone of influence (ZOI) of the MBH all stars are initially set up to have the same binding energy as a star on a circular orbit at the zone of influence ($E_{ZOI}$). A spherical cusp potential of the form of a harmonic oscillator is added with the frequency chosen to give the same zone of influence for the MBH as a stellar velocity dispersion of 200 km/s. 

% disk model
The disk model is an accretion $\alpha$-disk as described in \cite{Goodman03} modified at $r = 500$ Schwarzschild radii ($R_S$) to fall off as $r^{-2}$ to limit the disk mass to approximately $2 \times 10^6 \msun$ inside $R_{ZOI}$. For comparison this region contains $10^{8}\msun$ of stars, therefore the disk mass can be neglected when integrating particle orbits.

% interaction
The interaction between the stars and the disk is modelled by a single velocity impulse when the particles cross the plane of the disk, at radial distance $r_d$, of the form given in \cite{JMK2005}.

% code
To model the particle motions we developed a special purpose parallel code that uses the Bulirsch-Stoer method for the numerical integration of the orbits. This code has more than sufficient accuracy to ensure that stars cannot migrate into the loss cone of the disk by numerical errors. 

% end states
The possible final states for a star are that it is absorbed into the disk, it is directly captured by the MBH, or it remains in an uncaptured orbit. Stars absorbed into the disk are assumed to eventually be destroyed and their material accreted onto the MBH. The absorption of a star is said to occur when the orbital angular momentum of the star being aligned to the disk while the apocentre is within the disk. A capture occurs whenever the star passes within $4 R_S$ of the MBH (the condition for capture in a Schwarzschild black hole when the Newtonian orbit is extrapolated, in the limit of high eccentricity). 
This is often due to the relative inclination at the first plunge being high which leads to the star being brought close to the MBH before the angular momentum has time to align with the disk.%, as seen inFigure 2.

\begin{figure}
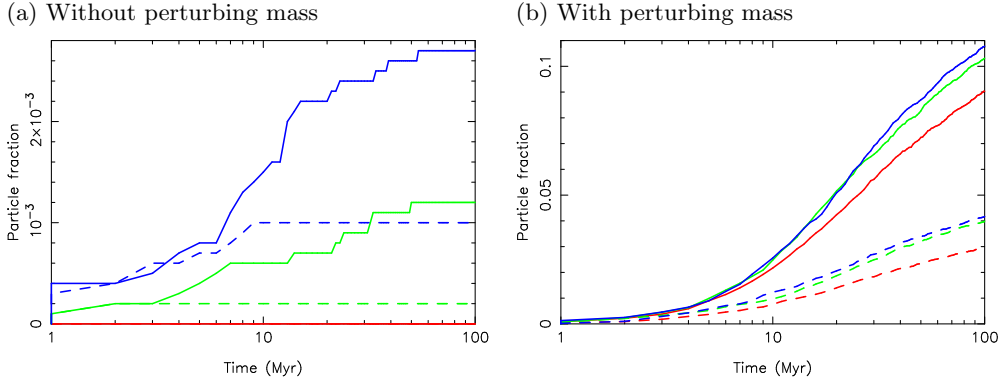

\begin{centering}$\begin{array}{cccc}
\multicolumn{1}{l}{\mbox{(a) Without perturbing mass}} & 
\multicolumn{1}{l}{\mbox{(b) With perturbing mass}}\\[-0.1cm]
\includegraphics[width=\figsizeFour,angle=270]{DT_FullRuns_log.ps} & 
\includegraphics[width=\figsizeFour,angle=270]{DT_RPe0_log.ps}
\end{array}$ \par\end{centering}
\caption{Left panel shows the fraction of particles captured by the MBH directly ({\it dashed}) and via absorption by the disk ({\it solid}) for the three stellar cusp potentials. Right panel shows the fractions for the same initial conditions but with a perturbing mass of $10^6 \msun$ at the zone of influence.}
\label{fig1} 
\end{figure}

% describe results
A sample of results from \cite{Kennedy_etal10} is shown in Figure~\ref{fig1}. For the cases presented here $10^4$ test particles were integrated, all of them with a binding energy equivalent to a circular orbit at the zone of influence for the MBH in the assumed stellar cusp. We explore three models for the gravitational potential of the stellar cusp: spherically symmetric ({\it red}), oblate ({\it green}) and triaxial ({\it blue}). Panel (a) shows the fraction of stars absorbed into the disk ({\it solid lines}) or directly captured by the MBH ({\it dashed lines}). The case with no perturbing masses added to the fixed potential is shown in panel (a), while panel (b) shows the effect of adding a $10^6 \msun$ point mass on a circular orbit at the zone of influence in the plane of the gas disk. Similar results are also achieved for perturbing masses on inclined and/or elliptical orbits. 

\section{Conclusions and Implications}

We find that the addition of a $10^6 \msun$ perturbing mass greatly enhances the rate at which the loss cone of the disk is replenished. In fact the case with the perturber shown in Figure~\ref{fig1} (b) is very close to the most efficient full loss cone case (not shown). In addition, approximately twice as many stars that are brought into the loss cone of the disk will end up being absorbed into the disk rather than directly captured by the MBH independent of the presence of a perturbing mass. The mass transferred into the disk from the stars is expected to be higher still since we have neglected mass loss of the stars due to stellar winds or stripping of the envelope as they pass through the disk. In summary we find this method very promising to bring material into the accretion disk by maintaining a full loss cone for absorption of stars in the accretion disk. The full details of this work will appear in an upcoming publication \cite{Kennedy_etal10}.

\end{document}